# EL IMPACTO DEL RUIDO BLANCO BINAURAL CON OSCILACIONES DE 100 A 750HZ EN LA MEMORIA DE TRABAJO VISUAL A CORTO PLAZO Y LA REACTIVIDAD DE ONDAS CEREBRALES ALFA Y BETA

Cesar Salas Guerra

Universitat Autònoma de Barcelona

P.O Box 931

Hatillo, PR 00659, USA

cesar.salasg@e-campus.uab.cat

## RESUMEN

De acuerdo algunos investigadores el ruido es concebido típicamente como factor perjudicial en el desempeño cognitivo afectando la percepción, toma de decisiones y la función motora. No obstante, en estudios recientes se asocia al ruido blanco con la concentración y la calma, por lo tanto, esta investigación busca establecer el impacto del ruido blanco binaural en el desempeño de la memoria de trabajo y visual a corto plazo, la actividad cerebral alfa – beta y la atención – meditación, mediante el uso de dos estímulos auditivos con rangos de frecuencia de (100 a 450hz) y (100 a 750hz). Este estudio se realizó en la ciudad de Montes Claros, República de Brasil, donde se evaluó a siete participantes (n = 7) con una edad promedio de 36.71±, y dos grupos de edad (GP1) 21 a 30 y (GP2) 41 a 50 de escolaridad media a universitaria. Dentro del proceso experimental se realizaron pruebas de memoria visual a corto plazo mediante el uso de la batería de evaluación cognitiva general CAB de CogniFit™, así como el registro de actividades cerebrales mediante el uso de Electroencefalograma monopolar y los algoritmos eSense™. Con los resultados obtenidos y mediante el uso de pruebas estadísticas podemos inferir que el ruido blanco binaural con oscilaciones de 100 a 750 Hz contribuyeron con el rendimiento de la memoria visual de trabajo a corto plazo.

## Palabras Claves

Brain Computer Interface. Estimulación Acústica. Memoria Corto Plazo. Neurorretroalimentación. Ondas Cerebrales. Procesos Mentales.

## 1. INTRODUCCION

Las investigaciones basadas en BCI o 'Brain Computer Interface"; desde su inicio estuvieron relacionadas a las funciones musculares y nerviosas (1), no obstante, en estos últimos años se han desarrollado nuevos paradigmas basados en la relación "humano-información-maquina" (2); y modernas técnicas no invasivas como la neurorretroalimentación (3), que le permite al cerebro desarrollar procesos de aprendizaje homeostático con oscilaciones flexibles entre estados de activación menores y mayores, lo cual ha contribuido en el desarrollo de nuevas investigaciones en Neurotecnología, con implicaciones en procesos de interacción humano-maquina (4).

El ruido es concebido típicamente como factor perjudicial en el desempeño cognitivo (5), con ciertas características inherentes en el procesamiento neural que afectan la percepción, toma de decisiones y función motora (6). No obstante, existen diferentes clases de ruido (7), entre ellos está el "Ruido Blanco" el cual de acuerdo con numerosos estudios tiene cierta relación con el aprendizaje y la calma.

A la memoria prospectiva se lo define (8) como una habilidad para recordar y llevar a cabo ciertas operaciones en el futuro, esto conlleva un proceso de recuperación de información almacenada (9); demostrando así su eficiencia en los procesos de codificación e interpretación, ya que los recuerdos a corto plazo (10) desarrollan suficientes recursos de recuperación semántica. Empero, la memoria icónica exige y requiere atención (11) entendiéndose así la relación de este tipo de memoria con la atención.

## 2. OBJETIVOS

Existe un creciente número de investigaciones que establecen el impacto del ruido en los sistemas neuronales (12), sin embargo, el ruido blanco se asocia con la concentración y la calma (13) por tal motivo este estudio tiene como objetivo establecer el impacto significativo de las oscilaciones de ruido blanco binaural de 100 a 750hz, en la memoria visual a corto plazo, la actividad cerebral alfa – beta y la atención – meditación.

Por lo tanto, con los sujetos de estudio se realizaron pruebas de memoria visual a corto plazo mediante el uso de la batería de evaluación cognitiva general CAB de CogniFit™, registrándose las actividades cerebrales mediante el uso de Electroencefalograma monopolar y los algoritmos eSense™.

## 3. SUJETOS Y METODOS

### 3.1 Participantes

Este estudio se realizó en la ciudad de Montes Claros, estado de Minas Gerais en la República Federal de Brasil en enero del 2018, la cual se desarrolló bajo un proceso de estimulación acústica binaural, mediante el registro electroencefalográfico monopolar tomado de la región frontopolar Fp1 del hemisferio izquierdo, utilizando dos niveles de estímulo acústico con frecuencias distintas (100 a 450hz) y (100 a 750hz) llamados en este estudio distractores de atención ambiental.

En el estudio se evaluaron 7 participantes (n = 7) con una edad promedio de 36.71±, y organizados en dos grupos (GP1 ) 21 a 30 y (GP2) 41 a 50, con escolaridad media a universitaria, el reclutamiento se realizó de forma incidental, mediante la difusión del proyecto, de viva voz a través de la red social LinkedIn, la participación fue siempre voluntaria, no remunerada y con el requisito previo de la lectura y aprobación de la hoja de información del proyecto de investigación y la firma de su respectivo consentimiento informado. Los participantes debían cumplir con los criterios de inclusión que se detalla a continuación:

a) Residentes en la ciudad de Montes Claros, en el Estados de Minas Gerais en la República Federal de Brasil.

b) Su primera lengua debe ser el portugués.

c) Los rangos de edad son de 21 a 30 y de 41 a 50 Años.

d) Género 6 femenino y 1 masculino.

e) Escolaridad enseñanza media a universitaria.

f) La audición, visión y condiciones físicas deben ser adecuadas para realizar las evaluaciones (deben utilizarse las medidas protésicas correctoras, tales como el uso de gafas, audífono o cualquier otro dispositivo).

g) Capacidad suficiente de ver, escuchar y capacidad de usar cualquiera de sus dos extremidades para el uso de un dispositivo electrónico táctil.

Para los criterios de exclusión se consideró los siguientes factores:

a) Falta de voluntad o incapacidad del participante para colaborar adecuadamente en el estudio.

b) Cualquier patología del sistema nervioso central, que pueda afectar a la cognición como la enfermedad de: parkinson, huntington, tumor cerebral, hidrocefalia, parálisis supra nuclear progresiva, epilepsia, hematoma subdural, esclerosis múltiple, historia de infarto cerebral.

c) Episodio depresivo mayor o trastorno distímico, según los criterios DSM-V.

d) Enfermedad cardiovascular inestable o clínicamente significativa en los 6 meses anteriores y que, a juicio del clínico, pueda tener impacto en las capacidades mentales.

e) Diabetes insulinodependiente.

f) Historial o presencia de abuso de alcohol o drogas conocidas en los 24 meses anteriores a la elaboración de la prueba.

g) Cualquier situación que pudiera hacer al participante, según la opinión del investigador principal, inadecuados para el estudio.

## 3.2  Recolección de Datos
Dentro de este proceso de registro de datos macrofisiológico de estudio no invasivo, los participantes fueron instruidos a mantenerse tranquilos, y colaborar con el uso del EEG o Electroencefalograma monopolar, denominado también como BCI "Brain computer interface", el cual consta de un adaptador craneal externo ubicado en el lado izquierdo de la frente llamado en la nomenclatura de posición de electrodos como región frontopolar Fp1 del hemisferio izquierdo (14), así como el uso de audífonos de alta fidelidad que bloquearan sonidos externos buscando escuchar solamente las muestras de estímulo auditivo.

La Electroencefalografía (16) es un registro grafico de la actividad eléctrica cerebral, compuesto por un numero variado de ondas que se presentan de forma aislada o en grupos, es importante señalar que en diferentes estudios realizados (17) mediante la colocación de un sensor en la zona próxima al lóbulo frontal encargado de la inteligencia, memoria y personalidad permitirá estudiar el cambio neuronal cuando el participante se someta a un ruido que estimule su proceso cognitivo.

El dispositivo de recolección de datos dispone también de unos algoritmos llamados eSense ™ (15) los cuales miden los niveles de atención y meditación, estos algoritmos desarrollan procesos y espectros dinámicos de oscilación mediante adaptación de fluctuaciones naturales en base a tendencias de cada participante mediante el uso de distractores de atención ambiental. La escala asignada por el algoritmo eSense™ para atención y meditación es: desempeño bajo (1-39), desempeño normal o base (40-60), desempeño superior (61-100).

## 3.3  Estimulo Auditivo
La muestra de ruido blanco para el respectivo experimento estuvo compuesta por dos archivos llamados sample-1-sound.wav (100 a 450hz) y sample-2-sound.wav (100 a 750hz), las cuales se denomino distractores de atención ambiental. Para conformar el estímulo auditivo en las dos pruebas por cada participante se utilizó como fondo ambiental individual estos distractores por medio de audífonos de alta fidelidad, minimizando de esta manera la densidad y elasticidad del medio de propagación.

## 3.4  Prueba de Memoria
En el proceso de interacción biocibernética el participante usará la batería de evaluación cognitiva general CAB de CogniFit™, la cual asido probada en estudios anteriores (18) permitiendo medir la memoria visual de trabajo a corto plazo mediante un ejercicio online, esta prueba está inspirada en: Wechsler Memory Scale (WMS), Test Clásico Memory Malingering (TOMM), y Torre de Londres (TOL).

Las instrucciones entregadas a los participantes del experimento fue estar atentos a los estímulos que aparecerán en la pantalla posicionados y distribuidos de manera aleatoria, y siguiendo un orden, los estímulos irán iluminándose uno detrás del otro hasta completar una serie. Por lo que se deberá observar atentamente mientras es el turno del ordenador y el del participante. Ya que el usuario deberá recordar el orden de la presentación de los estímulos para posteriormente ir reproduciéndoles según se hayan presentado en su orden correspondiente. Luego de terminado la secuencia del ordenador el participante deberá reproducir exactamente en la misma orden de la presentación del ordenador.

El objetivo de esta prueba fue recordar la cantidad máxima de series y su respectiva secuencia de la manera más precisa posible y atender de manera constante los estímulos visibles (Flores) que se presentaban en la pantalla del computador. Juntamente con estas pruebas se usaron dos estímulos audibles, la evaluación de esta prueba estuvo compuesta por una escala de resultados del 1 al 36, realizándose una sola vez por estimulo en ambientes completamente similares.

## 3.5  Análisis de Datos
Los datos fueron analizados en el programa para MS Windows 10, Minitab® versión 18.1, donde se realizaron estadísticas descriptivas e inferenciales, con pruebas paramétricas y no paramétricas.

Este estudio consta de un total de siete variables explicadas a continuación:

1. Ruido blanco binaural (MCR1: 100-400Hz)
2. Ruido blanco binaural (MCR2: 100-750Hz)

3. Memoria visual de trabajo a corto plazo (PRU1-2)
4. Atención eSense™ (RAeS1-2)
5. Meditación eSense™ (RMeS1-2)
6. Ondas cerebrales alfa (ROA1-2)
7. Ondas cerebrales beta (ROB1-2)

DSD1: Datos demográficos - Edad de los participantes establecidos en dos grupos de edad.

PRU1: Prueba 1 - Resultados obtenidos en la prueba de memoria usando test de memoria CAB y distractor de atención ambiental MCR-1.

PRU2: Prueba 2 - Resultados obtenidos en la prueba de memoria usando test de memoria CAB y distractor de atención ambiental MCR-2.

RAeS1: Atención eSense™ 1 - Resultados de los registros de Atención obtenidos mediante el uso del EEG con eSense™, usando el distractor de atención ambiental MCR-1.

RAeS2: Atención eSense™ 2 - Resultados de los registros de Atención obtenidos mediante el uso del EEG con eSense™, usando el distractor de atención ambiental MCR-2.

RMeS1: Meditación eSense™1 - Resultados de los registros de Meditación obtenidos mediante el uso del EEG con eSense™, usando el distractor de atención ambiental MCR-1.

RMeS2: Meditación eSense™ 2 - Resultados de los registros de Meditación obtenidos mediante el uso del EEG con eSense™, usando el distractor de atención ambiental MCR-2.

ROA1: Alfa 1 - Resultados de los registros obtenidos en la adquisición de las ondas Alfa 1, usando el distractor de atención ambiental MCR-1.

ROA2: Alfa 2 - Resultados de los registros obtenidos en la adquisición de las ondas Alfa 2, usando el distractor de atención ambiental MCR-2.

ROB1: Beta 1 - Resultados de los registros obtenidos en la adquisición de las ondas Beta 1, usando el distractor de atención ambiental MCR-1.

ROB2: Beta 2 - Resultados de los registros obtenidos en la adquisición de las ondas Beta 2, usando el distractor de atención ambiental MCR-2.

Las mismas fueron sometidas al análisis estadístico correspondiente, para establecer un proceso ordenado y sistemático en el análisis e interpretación de datos se divide el estudio en experimentos y etapas de prueba, lo cual nos permitirá tener un orden y comprensión de las pruebas realizadas en los diferentes experimentos.

## 3.6 Primer Experimento

*Primera Etapa:* se realizó un análisis estadístico descriptivo de la muestra DSD1, y las variables, PRU1 y PRU2 (Anexo I, Tabla I).

*Segunda Etapa:* para establecer el supuesto de normalidad se usó la prueba de Anderson-Darling (19) (9) (Anexo I, Tabla II) la cual se basa en la función de la distribución empírica para establecer los supuestos de normalidad. Buscando realizar una prueba comparativa "post hoc", se realizó el mismo ensayo con la prueba Shapiro-Wilk (20), (21), (9), (Anexo I, Tabla II).

*Tercera Etapa:* con los datos de ubicación obtenido por las pruebas normalidad, se procede a realizar la prueba T estudiante (22) (Anexo I, Tabla III), para establecer la diferencia significativa entre las variables PRU1 y PRU2, y de esta forma determinar la existencia de diferencias significativas entre los resultados de estas bajo la influencia de los distractores de atención ambiental.

## 3.7 Segundo Experimento

*Primera Etapa:* se realizó un análisis estadístico descriptivo de las variables atención eSense™ (RAeS1-2) y meditación eSense™ (RMeS1-2) (Anexo I, Tabla I).

*Segunda Etapa:* para establecer el supuesto de normalidad se usó la prueba de Anderson-Darling (19) (9) (Anexo I, Tabla II) la cual se basa en la función de la distribución empírica para establecer los supuestos de normalidad. Buscando realizar una prueba comparativa "post hoc", se realizó el mismo ensayo con la prueba Shapiro-Wilk (20), (21), (9) (Anexo I, Tabla II).

*Tercera Etapa:* con los datos de ubicación obtenidos por las pruebas de normalización, se procede a realizar la prueba T estudiante (22) (Anexo I, Tabla III), para establecer la diferencia significativa entre las variables RAeS1-2 y RMeS1-2, determinando de esta esta forma si existe diferencias significativas entre los resultados de estas bajo la influencia de los distractores de atención ambiental.

## 3.8 Tercer Experimento

*Primera Etapa:* se realizó un análisis estadístico descriptivo de las variables alfa 1 y 2 (ROA1-2) y beta 1 y 2 (ROB1-2) (Anexo I, Tabla I).

*Segunda Etapa:* para establecer el supuesto de normalidad de la variable ROA1-2 y ROB1-2, se usó la prueba de Anderson-Darling (19) (9) (Anexo I, Tabla II), la cual se basa en la función de la distribución empírica para establecer los supuestos de normalidad. Buscando realizar una prueba comparativa "post hoc", se realizó el mismo ensayo con la prueba Shapiro-Wilk (20), (21), (9) (Anexo I, Tabla II).

*Tercera Etapa:* con los datos de ubicación obtenido por las pruebas de normalidad, y al hallarse que no cumplen con los requisitos de significancia se procede a realizar la prueba no paramétrica Mann-Withney (22), (20), (21), (9) (Anexo I, Tabla III). para establecer la diferencia significativa entre las variables ROA1-2 y ROB1-2.

## 4. RESULTADOS

*Primero*: Los resultados del estudio confirman en (experimento 1), que existe diferencia significativa entre la variable PRU1 (Me = <18.14) y PRU2 (Me = >22.43), con un valor de significancia (p value = 0.048), esta prueba T Student se realizó con un nivel de confianza del (95%). Estos hallazgos nos indican que existió un grado más alto de memorización visual a corto plazo en la PRU2, mediante el uso del distractor de atención ambiental MCR2, que en la PRU1 y la variable MCR1.

*Segundo:* Los hallazgos del (experimento 2), concluyen que no existió diferencia significativa entre la variable RAeS1 (Me =

<52.44) y RAeS2 (Me = >52.89), con un valor de significancia (p value = 0.408), esta prueba T Student se realizó con un nivel de confianza del (95%) por consiguiente, estos hallazgos nos indican que no existió algún tipo de impacto o influencia del distractor de atención ambiental MCR1-2, en las variables RAeS1-2.

*Tercero:* Los hallazgos de la segunda parte del (experimento 2), concluyen que no existe diferencia significativa entre la variable RMeS1 (Me = <57.13) y RMeS2 (Me = >57.92), con un valor de significancia (p value = 0.331), esta prueba T Student se realizó con un nivel de confianza del (95%) por consiguiente, estos hallazgos nos indican que no existió algún tipo de impacto o influencia del distractor de atención ambiental MCR1-2, en las variables RMeS1-2.

*Cuarto:* Los hallazgos del (experimento 3), concluyen que no existe diferencia significativa entre la variable ROA1 (Md = <11763.5) y ROA2 (Md = >12989.5), con un valor de significancia (p value = 0.346), esta prueba no paramétrica Mann-Whitney se realizó con un nivel de confianza del (95.44%) por consiguiente, estos hallazgos nos indican que no existió algún tipo de impacto o influencia del distractor de atención ambiental MCR1-2, en las variables ROA1-2.

*Quinto:* Los hallazgos del (experimento 3), concluyen que no existe diferencia significativa entre la variable ROB1 (Md = <11763.5) y ROB2 (Md = >12989.5), con un valor de significancia (p value = 0.421), esta prueba no paramétrica Mann-Whitney se realizó con un nivel de confianza del (95.44%) por consiguiente, estos hallazgos nos indican que no existió algún tipo de impacto o influencia del distractor de atención ambiental MCR1-2, en las variables ROB1-2.

## 5. DISCUSION

Los resultados de esta esta investigación nos permite observar que al ruido blanco no se le debe concebir únicamente como factor perjudicial en el desempeño cognitivo (5), ni tampoco como una característica inherente del procesamiento neural que afecta la percepción, la toma de decisiones y la función motora (6), con los resultados de este estudio experimental podemos inferir que el ruido blanco binaural con oscilaciones de 100 a 750 Hz contribuyen con el rendimiento de la memoria visual de trabajo a corto plazo.

Esta investigación es un componente dentro del estudio de las variables (Human Data Sensor) y sus operadores (sAuditivos - sVisuales – sTáctiles) como parte de investigaciones en curso, no obstante, se realizará a futuro un segundo estudio buscando observar los mismos resultados en la misma población de estudio, pero una muestra de pacientes de Alzheimer. Dentro de las limitaciones está el tiempo y los recursos con los cuales se hubiera podido trabajar con una muestra más grande.

## ANEXO I

**Tabla I.** *Estadística descriptiva de condiciones experimentales*

|          | EDAD   | PRU 1 | PRU 2 | RAeS 1 | RAeS 2 | RMeS 1 | RMeS 2 | ROA 1    | ROA 2     | ROB 1    | ROB 2     |
|----------|--------|-------|-------|--------|--------|--------|--------|----------|-----------|----------|-----------|
| MUESTRA  | 7      | 7     | 7     | 7      | 7      | 7      | 7      | 14       | 14        | 14       | 14        |
| MEDIA    | 36.71  | 18.14 | 22.43 | 52.44  | 52.89  | 57.13  | 57.92  | 19985    | 23600     | 13245    | 17559     |
| ST-DEV   | 11.18  | 8.99  | 8.98  | 8.03   | 5.88   | 7.48   | 5.54   | 8083     | 10442     | 5028     | 10140     |
| VARIANZA | 124.90 | 80.81 | 80.82 | 64.46  | 34.58  | 55.99  | 30.71  | 65335610 | 109031467 | 25283300 | 102809606 |
| COEF-VAR | 30.44  | 49.55 | 40.03 | 15.31  | 11.12  | 13.10  | 9.57   | 40.45    | 44.25     | 37.96    | 57.75     |
| MINIMO   | 21     | 7     | 7     | 37.88  | 42.84  | 45.39  | 51.89  | 8029     | 11935     | 7139     | 8321      |
| Q1       | 26     | 13    | 17    | 46.35  | 49.20  | 49.09  | 51.96  | 14567    | 15810     | 9991     | 10299     |
| MEDIANA  | 40     | 17    | 24    | 55.91  | 54.03  | 59.70  | 60.69  | 17246    | 20072     | 11764    | 12990     |
| Q3       | 48     | 21    | 27    | 56.23  | 55.39  | 61.49  | 63.11  | 24779    | 36700     | 15284    | 29006     |
| MAXIMO   | 50     | 36    | 36    | 62.37  | 62.05  | 66.57  | 64.39  | 37879    | 41886     | 24727    | 17588     |

**Tabla II.** *Pruebas de normalidad de condiciones experimentales*

| Test     |         | PRU 1 | PRU 2 | RAeS 1 | RAeS 2 | RMeS 1 | RMeS 2 | ROA 1 | ROA 2 | ROB 1 | ROB 2 |
|----------|---------|-------|-------|--------|--------|--------|--------|-------|-------|-------|-------|
|          | MUESTRA | 7     | 7     | 7      | 7      | 7      | 7      | 14    | 14    | 14    | 14    |
| Anderson | MEDIA   | 18.14 | 22.43 | 52.44  | 52.89  | 57.13  | 57.92  | 19985 | 23600 | 13245 | 17559 |
| Darling  | ST-DEV  | 8.898 | 8.979 | 8.029  | 5.881  | 7.483  | 5.542  | 8083  | 10442 | 5028  | 10140 |
|          | AD      | 0.489 | 0.228 | 0.041  | 0.306  | 0.316  | 0.595  | 0.502 | 1.089 | 0.765 | 1.457 |
|          | P-VALUE | 0.136 | 0.703 | 0.257  | 0.465  | 0.438  | 0.073  | 0.172 | 0.005 | 0.035 | 0.005 |
|          | MUESTRA | 7     | 7     | 7      | 7      | 7      | 7      | 14    | 14    | 14    | 14    |
| Shapiro  | MEDIA   | 18.14 | 22.43 | 52.44  | 52.89  | 52.89  | 57.92  | 19985 | 23600 | 13245 | 17559 |
| Wilk     | ST-DEV  | 8.989 | 8.979 | 8.029  | 5.881  | 5.881  | 5.542  | 8083  | 10442 | 5028  | 10140 |
|          | RJ      | 0.929 | 0.976 | 0.898  | 0.964  | 0.964  | 0.923  | 0.960 | 0.919 | 0.931 | 0.886 |
|          | P-VALUE | 0.100 | 0.100 | 0.100  | 0.100  | 0.100  | 0.100  | 0.100 | 0.027 | 0.044 | 0.010 |

**Tabla III.** *Análisis con Prueba T Student y Mann-Whitney.*

| Test | | PRU 1 - 2 | RAeS 1 - 2 | RMeS 1 - 2 | ROA 1 - 2 | ROB 1 |
|------|--|-----------|------------|------------|-----------|-------|
| | DIFERENCIA | -4.29 | -0.45 | -0.79 | 14 | 14 |
| *Prueba* | LS-95% | 4.18 | 6.37 | 5.53 | 19985 | 13245 |
| *T Student* | METODO | -- | -- | -- | 8083 | 5028 |
| | $\mu_1$: | $M_e$ muestra 1 | $M_e$ muestra 1 | $M_e$ muestra 1 | 0.502 | 0.765 |
| | $\mu_1$: | $M_e$ muestra 2 | $M_e$ muestra 2 | $M_e$ muestra 2 | 0.172 | 0.035 |
| | DIFERENCIA | $\mu_1 - \mu_2$ | $\mu_1 - \mu_2$ | $\mu_1 - \mu_2$ | | |
| | H-NULA | $H_0: \mu_1 - \mu_2 = 4.29$ | $H_0: \mu_1 - \mu_2 = 0.45$ | $H_0: \mu_1 - \mu_2 = 0.79$ | 14 | 14 |
| | H-ALTERNA | $H_1: \mu_1 - \mu_2 < 4.29$ | $H_1: \mu_1 - \mu_2 < 0.45$ | $H_1: \mu_1 - \mu_2 < 0.79$ | 19985 | 13245 |
| | T VALUE | -1.80 | -0.24 | -0.45 | 8083 | 5028 |
| | P VALUE | 0.048 | 0.408 | 0.331 | 0.960 | 0.931 |

*Análisis con Prueba Mann-Whitney*

| Test | | ROA 1 - 2 | ROA 1 - 2 |
|------|--|-----------|-----------|
| | Mediana Alfa 1 \| Beta 1 | 17206 | 11763.5 |
| | Mediana Alfa 2 \| Beta 2 | 20072 | 12989.5 |
| | Diferencia $\eta_1 - \eta_2$ | -2539 | -1512.9 |
| | IC para la diferencia | -8996, 2946 | -7105.3, 1963.5 |
| *Prueba* | Confianza lograda | 95.44% | 95.44% |
| *Mann-Whitney* | Método | -- | -- |
| | $\eta_1$: | mediana de Alfa 1 | mediana de Beta 1 |
| | $\eta_2$: | mediana de Alfa 2 | mediana de Beta 2 |
| | DIFERENCIA | $\eta_1 - \eta_2$ | $\eta_1 - \eta_2$ |
| | Hipótesis Nula | $H_0: \eta_1 - \eta_2 = 0$ | $H_0: \eta_1 - \eta_2 = 0$ |
| | Hipótesis Alterna | $H_1: \eta_1 - \eta_2 \neq 0$ | $H_1: \eta_1 - \eta_2 \neq 0$ |
| | Valor W | 182 | |
| | Valor p | 0.346 | |